\documentclass[pdftex,onecolumn,epjc3]{svjour3}
\RequirePackage[T1]{fontenc}
\RequirePackage{fix-cm}
\RequirePackage{graphicx}
\RequirePackage{mathptmx} 
\RequirePackage{latexsym}
\RequirePackage{flushend}
\RequirePackage[numbers,sort&compress]{natbib}
\RequirePackage[colorlinks,citecolor=blue,urlcolor=blue,linkcolor=blue]{hyperref}
\RequirePackage{amsmath}

\newcommand{\ben}{\begin{eqnarray}}
\newcommand{\een}{\end{eqnarray}}
\newcommand{\be}{\begin{equation}}
\newcommand{\ee}{\end{equation}}
\newcommand{\n}{\label}
\newcommand{\no}{\noindent}

\newcommand{\ro}{\rho}
\newcommand{\om}{\omega}

\journalname{Eur. Phys. J. C}

\begin{document}
\title{ Holographik, the k-essential approach to interactive models with modified holographic Ricci dark energy}

\author{M\'onica Forte\thanksref{e1,addr1}}
\thankstext{e1}{e-mail:forte.monica@gmail.com}
\institute{Departamento de F\'isica, Facultad de ciencias Exactas y Naturales, Universidad de Buenos Aires, 1428 Buenos Aires, Argentina \label{addr1}}
\date{Received: date / Accepted: date}
\maketitle
\begin{abstract}
We make a scalar representation of interactive models with cold dark matter and modified holographic Ricci dark energy through unified models driven by scalar fields with non-canonical kinetic term.  These models are applications of the formalism of exotic k-essences   generated by the global description of cosmological models with two interactive fluids in the dark sector and in these cases they correspond to usual k-essences. The formalism is applied to the cases of constant potential in Friedmann-Robertson-Walker geometries.
\end{abstract}


\section{Introduction}
\label{intro}

There are a number of cosmological observations, particularly from Type Ia Supernovae \cite{Riess:1998cb,Perlmutter:1998np,Amanullah:2010vv}, Cosmic Microwave Background Radiation
\cite{Spergel:2006hy}, and Baryon Accoustic Oscillation \cite{Weinberg:2012es,Percival:2007yw} showing an accelerating effect on the expansion of our universe.  Therefore, there must be a cosmological component responsible for the repulsive behavior that allows counteract and overcome the gravitational attraction. For this constituent with negative pressure, dubbed dark energy (DE), there have been some proposals. The cosmological constant seems to give the best fit with the observations but also there are good dynamical models including quintessence \cite{Zlatev:1998tr,Caldwell:1997ii,Frieman:1995pm,Peebles:1987ek,Ratra:1987rm}, k-essence \cite{ArmendarizPicon:2000dh,ArmendarizPicon:2000ah,Copeland:2006wr,Chimento:2003ta,Chiba:2002mw,Tsujikawa:2004dp,Malquarti:2003nn,Aguirregabiria:2004te,Chimento:2004jm,Chimento:2005ua,ArmendarizPicon:2005nz,Li:2002wd}, models with internal structure as quintom \cite{Guo:2004fq,Cai:2009zp,Feng:2004ad,Feng:2004ff} and N-quintom \cite{Chimento:2008ws} and applications of holographic principle \cite{'tHooft:1993gx} to cosmology \cite{Fischler:1998st,Aharony:1999ti,Bousso:2002ju,Bousso:1999xy,Hsu:2004ri}. 
The other majority contribution to the source of Einstein equations is called dark matter (DM), and is the ingredient that comes to supplement the lack of observed non-relativistic matter. Again, we cannot say anything about its nature and moreover, we cannot argue with some symmetry or microphysical criteria that it is evolving regardless of the DE. In fact, the possibility of an interaction between DM and DE has received many attention in the literature \cite{Amendola:1999er,Mangano:2002gg,delCampo:2004wc,Farrar:2003uw,Guo:2004xx,Guo:2004vg,Cai:2004dk,Gumjudpai:2005ry,Curbelo:2005dh,Zimdahl:2001ar,Chimento:2003iea,Olivares:2005tb} and appears to be even favored over non-interacting cosmologies \cite{Szydlowski:2005ph}. 
This work has the goal of showing the connection between models led by common k-essences (but with special conditions on the signs of the first and second derivatives) and interactive models of cold dark matter (CDM) and modified holographic Ricci type dark energy (MHRDE) fluids. 
We nickname holographik to these common k-essences to stand out the fact that they are related to interactive models where the dark energy corresponds to a holographic fluid.
The idea has precedents in the linking of exotic quintessences \cite{Chimento:2007da,Forte:2016cwm} 
or exotic k-essences \cite{Forte:2015oma} with interactive systems of two arbitrary perfect fluids, but here the purely k-essence $\phi$ is derivable from a Lagrangian of the form $\mathcal{L}=-V_0F(\dot\phi^2)$ and the interactive systems are compound with fluids whose continuity equation can be replaced by a modified equation using constant coefficients. 
The MHRDE fluid used here was proposed in \cite{Granda:2008tm} as a particular class of the more general holographic Ricci type dark energy  introduced in \cite{Nojiri:2005pu} and it was the unique holographic component of a cosmological model that avoided the problem of causality \cite{Li:2004rb}. This statement can be explained as follows. According to the application of the holographic principle to cosmology, the vacuum  density of energy can be bounded by the full energy inside a region because it cannot exceed the mass of a black hole of the same size. From effective quantum field theory, an effective infrared (IR) cut-off can saturate the length scale that is included in the expression of the vacuum density of energy and in literature, the IR cut-off has been taken as the Hubble horizon, or the particle horizon, or the event horizon and also as some generalized IR cut-off. The papers devoted to holographic dark energy models with Hubble horizon or particle horizon as the IR cut-off, have shown that these models cannot lead to the current accelerated expansion of the universe. When event horizon is taken as the cut-off, as future event  horizon is a global concept of space-time while the density of dark energy is a local quantity, the relation between them will raise challenges to the causality.  These leads to the introduction of the holographic Ricci type  dark energy, where the IR cut-off is taken as proportional to the Ricci scalar curvature, where the problem can be avoided. In the context of interactive systems the MHRDE fluid was used in a plethora of models \cite{Chimento:2011dw,Chimento:2013se,P:2013cmq,Li:2014eba,Pasqua:2015bpm,Sharif:2012zza,delCampo:2011jp,Landim:2015hqa,Bamba:2012cp}.

The paper is organized as follows: In Section 2 we consider the non-canonical scalar representation of an interacting cosmological model realized with CDM and a MHRDE fluid and introduce the expressions of the different physical magnitudes in terms of the constant potential $V_0$ and of the suitable kinetic functions $F$ of a k-essence field $\phi$. In Section 3 we gain deeper insight into the subject analyzing the equation that must be fulfilled by the kinetic functions and the related interactions $Q(V_0, F)$. Also there, we show worked examples in both ways. On the one hand, for a given interaction we obtain the corresponding kinetic function and on the other hand we discover which interaction can be considered associated with widely studied k-essences. In Section 4 we draw conclusions about the examples in terms of the workability provided by the scalar representation and also on the generation of new functional forms of interaction that can be studied analytically.


\section{The holographic k-essence }


We consider a model consisting of two perfect fluids with an energy-momentum tensor $T_{ik} = T ^{(1)}_{ik}+ T ^{(2)}_{ik}$ where $T^{ (n)}_{ik}= (\rho_n + p_n)u_iu_k +p_n g_{ik}$ , being $\rho_n$ and $p_n$ the density of  energy and the equilibrium pressure of fluid $n$ and $u_i$ their four-velocity. 
Assuming that the two fluids interact between them in a spatially flat, homogeneous and isotropic Friedmann-Robertson-Walker (FRW) cosmological background, the Einstein equations reduce to:
\be
\n{1}
3H^2= \ro_1 + \ro_2\equiv \ro,
\ee
\vskip -0.2cm
\be
\n{2}
\dot\ro_1 + \dot\ro_2 +3H[(1+\om_1)\ro_1 + (1+\om_2)\ro_2]=\dot\ro+3H(1+\om)\ro=0,
\ee
\vskip 0.3cm
\no where $H = \dot a/a$ and $a$  stand for the Hubble expansion rate and the scale factor respectively and where we consider equations of state (EoS) $\om_i=(p_i/\ro_i)$ for $ i=1, 2$. 
Above, we have assumed an overall perfect fluid description with an effective equation of state, $\om = p/\ro = -2\dot H /3H^2 - 1$, where $p = p_1 + p_2$ and $\ro = \ro_1 +\ro_2$.  The dot means derivative with respect to the cosmological time and from equations (\ref{1}), (\ref{2}) we get
\be
\n{3}
-2\dot H  = (1 + \om_1)\ro_1 + (1+ \om_2)\ro_2 = (1 + \om)\ro.
\ee

In this paper a more general version of the holographic fluid described in \cite{Granda:2008dk,Granda:2008tm} is used as DE. This is the simplest case where the density of energy of the DE is expressed as a general function of Hubble parameter and its derivative, for which the models avoid the causality problem. Then, the holographic  density of energy $\rho_2$ is written as
\be
\n{4}
\ro_2^{MHRDE}=\frac{2}{A-B}(\dot H +\frac{3}{2}AH^2),
\ee

\no where $A$ and $B$  are two arbitrary constants that we can suppose that they satisfy  $A > B > 0$. From (\ref{3}) and (\ref{4}) we obtain 
\be
\n{5}
(1 + \om_1)\ro_1 + (1+ \om_2)\ro_2 = A \ro_1 + B \ro_2,
\ee
\no which is a very useful relation because in our description of the interactive system, the equations of state must be constant and in general it does not happen with the EoS of the holographic fluid. Note also that the expressions (\ref{1}), (\ref{2}) and (\ref{5}) allow us to write the partial densities of energy as

\be
\n{6}
\ro_1 = -\frac{B \ro +\ro'}{A-B}  \qquad  \ro_2 = \frac{A \ro +\ro'}{A-B},
\ee
for $\ro'= \dot\ro/3H$.

The interaction $Q$ that connects both fluids is specified through the partial equations of conservation

\begin{subequations}
\n{7}
\be
\n{7a}
\dot\ro_1 + 3H(1+\om_1)\ro_1= -3HQ
\ee
\be
\n{7b}
\dot\ro_2 +3H(1+\om_2)\ro_2=3HQ.
\ee
\end{subequations}

Or better, a modified interaction $Q_M$ can be defined by using the relations (\ref{5}) and (\ref{7}) by means of
\begin{subequations}
\n{8}
\be
\n{8a}
\dot\ro_1 + 3HA\ro_1= -3HQ_M
\ee
\be
\n{8b}
\dot\ro_2 +3HB\ro_2=3HQ_M.
\ee
\end{subequations}

 Clearly, the relation between $Q_M$ and $Q$ is $Q_M=Q+(1-A)\rho_1=Q+(B-\om_2-1)\rho_2$, where we apply the formalism to interactions between cold dark matter (CDM) and modified holographic Ricci type dark energy (MHRDE) fluid.

Now, as it  was done with the exotic canonical scalar field in  \cite{Chimento:2007da} and with the exotik field in \cite{Forte:2015oma},  we propose  that the interactive system as a whole be represented by a unified model driven by a special class of  purely k-essence field $\phi$ (labeled by a constant potential $V_0$ and a kinetic function $F(x)$, $ x=-{\dot\phi}^2$), through the relationship
\be
\n{9}
(1 + \om)\ro= A\ro_1 + B\ro_2= -2V_0xF_x(x), \qquad   F_x= \frac {d F(x)}{d x}.
\ee

Then, the global density of energy $\rho$ and  the global pressure $p=\om\ro$ can be written as 
\be
\n{10}
\ro = V_0(F(x)-2xF_x(x)), \qquad  p = -V_0F(x).
\ee

The field  $\phi$ satisfies the equation of movement
\be
\n{11}
\left[F_x+2xF_{xx}\right]\ddot\phi+3HF_x\dot\phi=0 \qquad F_{xx}=d F_x/d x,
\ee
that allows us to find the functional form of the k-field $ \phi $ once the kinetic function $F(x)$ is given. 
If the kinetic function is strictly monotonic $F_x \ne 0$, there is the well-known first integral 
\be
\n{12}
\sqrt{-x}F_x=m_0a^{-3},
\ee
\no for $ m_0 $ a constant  of integration. Alternatively, when the kinetic functions have an extreme $x_e=x(t_e)$ such that $F_x(x_e)=0$, the above first integral (\ref{12}) does not exist. Instead, at time $t=t_e$, the equation (\ref{11}) is reduced to $ x_e F_{xx} (x_e)\ddot\phi|_{t_e} = 0 $ and thus it must happen that $ \dot\phi $ has a root or an extreme at $t=t_e$, or that $F(x)$ has a saddle point at $x_e$. We will not address cases with non-monotonic kinetic functions.

We must note that, from (\ref{4}), (\ref{6}), (\ref{10}) and (\ref{11}) the partial densities of energy are
\begin{subequations}
\label{13}
\be
\label{13a}
\ro_1= -\frac{V_0}{A-B}(BF(x)-2xF_x(x)(B-1))
\ee
\be
\n{13b}
\ro_2^{MHRDE}= \frac{V_0}{A-B}(AF(x)-2xF_x(x)(A-1))=\frac{A-1-\om}{A-B}\ro
\ee
\end{subequations}

\no and  therefore, being $A-B>0$  the maximum possible value for the overall EoS should be  $\om=\om_{max}=A-1$. 
This one is the first characteristic that these "special" k essences must fulfill and interestingly, it comes exclusively from the associated interactive models using MHRDE fluids because $\ro_2^{MHRDE}$ must be non-negative. The expression for the global equation of state $\om$ in the unified representation of the k-essence, is
\be
\label{14}
\om= - \frac{ F}{F - 2xF_x},
\ee
\no and so (\ref{13b}) and (\ref{14}) imply $-2xF_x/(F - 2xF_x)\le A$.

Also, from (\ref{3}) and (\ref{5}) is $\om=(A-1+(B-1)r)/(1+r)$, where $ r= \ro_2/\ro_1$. Thus, if the universe supports a constant EoS $\om=\om_0$, then the ratio between densities must be a constant $ r = r_0 =(A-1-\om_0)/(1+\om_0-B)$. Conversely, in these models with interactive MHRDE fluid, we cannot have a stationary solution to the problem of coincidence, $r=r_0$, without paying the price of a universe with constant EoS $\om_0$. In that sense, from (\ref{14}) we can see that the polynomial kinetic functions $F=(-x)^n$ with $n=$ constant, have constant $\om= (2n-1)^{-1}$.  Therefore the interactive models with interactions associated with these $F$, should not be considered interesting examples to describe realistic cosmological models.

\begin{figure}[htbp]
\begin{center}
\includegraphics[height=8cm,width=10cm]{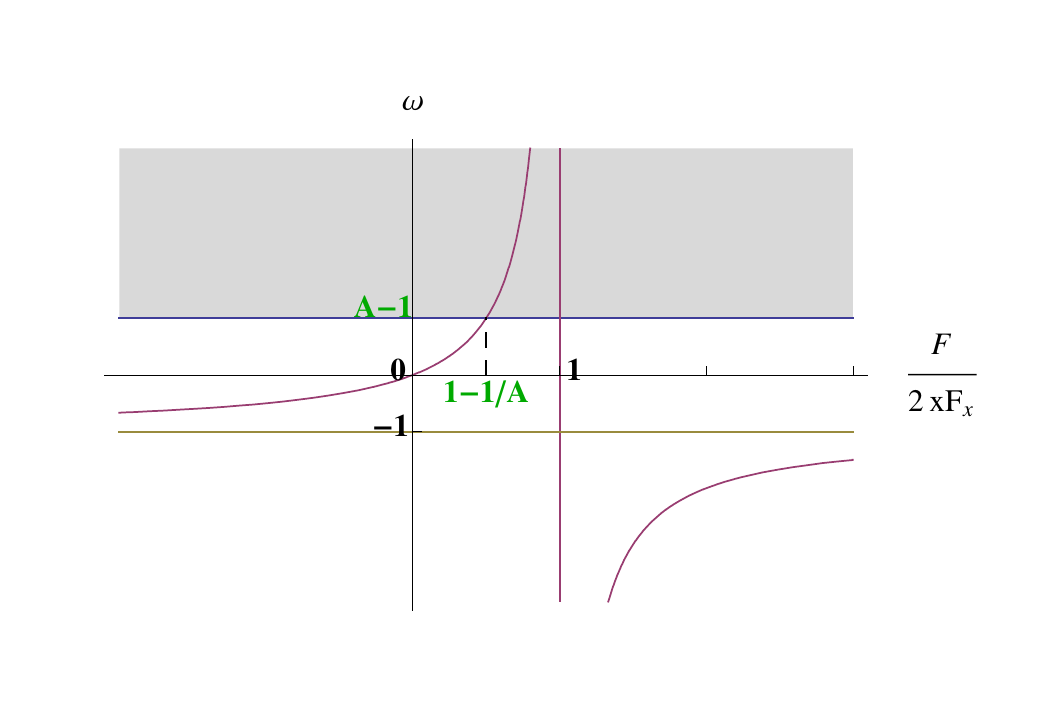}
\vskip-0.7cm
\caption{\label{fig:1}Evolution of the global equation of state for the holographik unified model as a function of the magnitude $ g=F/(2xF_x)$. The shaded area corresponds to the prohibited values $\om\ge A-1$, throughout all the evolution of the model. The maximum $\om_M=A-1$ is reached at $g=(A-1)/A$ belonging to the left branch of the graph, the more useful in modeling realistic universes. The right branch is related with phantom universes. The models with asymptotic stiff behavior must have $g \le 1/2$ and those with asymptotic radiation behavior must have $g \le 1/4$. The models with asymptotic dust behavior must have $g  \le 0$ and $F \ge 0$. }
\end{center}
\end{figure}

 The figure~\ref{fig:1} describes the global EoS $\om = g/(1-g)$ in terms of the auxiliary function $g \equiv F/(2xF_x)$ and also shows the prohibited zone $\om \ge A-1$. There, the left branch ($g < 1-1/A$) correctly describes a unified model which behavior interpolates between a stiff \cite{Zeldovich:1972zz,Chavanis:2014lra}, radiation or dust type (for $ A = 2,  4/3 $ or $ 1 $) and a cosmological constant type.  The right branch ($g > 1$) describes phantom models provided that the EoS is kept $\om < -1$ along the whole cosmological history. 

\begin{figure}[tbp]

\centering

\includegraphics[height=7cm,width=10cm]{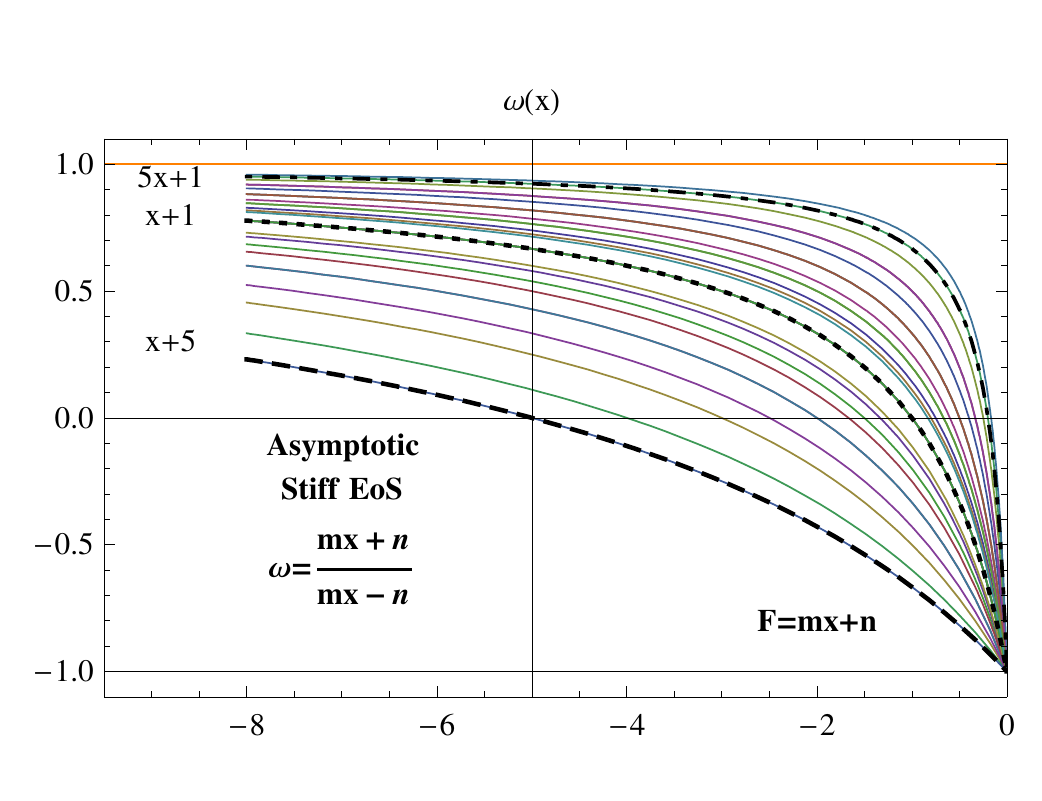}
\hfill

\includegraphics[height=7cm,width=10cm]{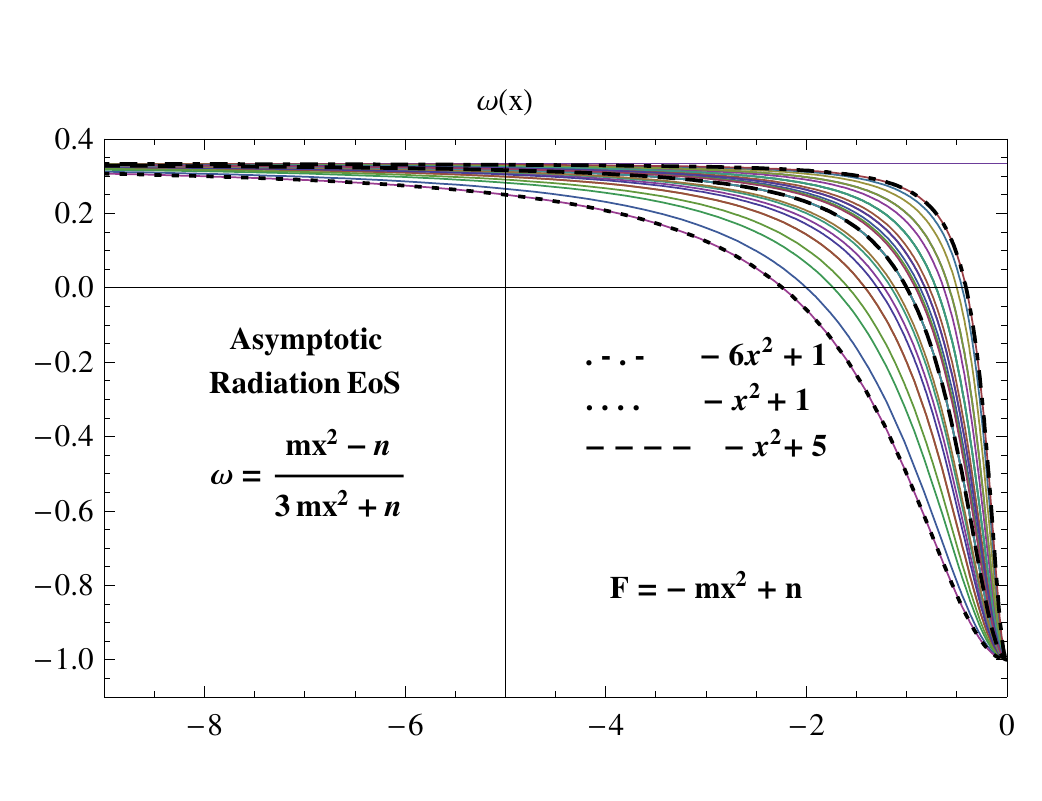}
\hfill

\includegraphics[height=7cm,width=12cm]{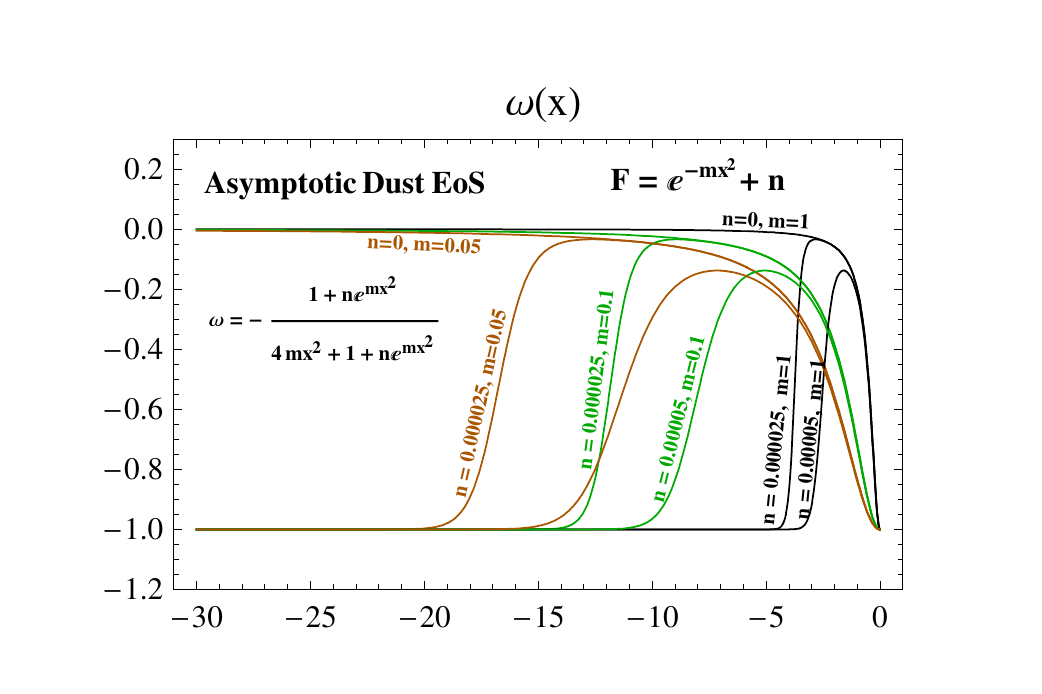}\hfill

\caption{\label{fig:2} Top panel: EoS interpolating between $\om= 1$ and $\om= -1$ corresponding to the kinetic functions $F[x]=  -mx+n$, $m=1,2,3,4,5,6$, $n=1,2,3,4,5$.Intermediate Panel: EoS interpolating between $\om= 1/3$ and $\om= -1$ corresponding to the kinetic functions $F[x]= -mx^2+n$, $m=1,2,3,4,5,6$, $n=1,2,3,4,5$. Bottom Panel: EoS interpolating between $\om= 0$ and $\om= -1$ or without dust like era but with an accelerated behavior at early time for tiny $n$, corresponding to the kinetic functions $F(x)=e^{-mx^2}+n$,  with $m = 1, 0.1, 0.05$, $n= 0, 0.00005, 0.05$.}

\end{figure}

Let us focus on left branch. The bound $\om \le A-1$ results in the bound $g \le 1-1/A$ and therefore in the "bounding" functions $F(x)_{max}= F_0(-x)^{A/(2(A-1))}$ for  $A=2$ or $A= 4/3 $ and anyone for which $g < 0$ if $A=1$. The meaning of "bounding" is evident in figure~\ref{fig:2}, where the general behaviors of $\om (x)$  for different functions $F$ appear "limited" by the curve with $n=0$

Other two conditions exist to carry out for these functions, which come from the reality of the Hubble factor $H$ and from the stability of the model. From (\ref{10}) the total density of energy can be written as $\ro/V_0=2xF_x(g-1)$. and with positive potentials and $g < 1$  it must always be observed that $ F_x>0 $. Therefore, this second condition leads to $F<0$ for $0<g<1-1/A$ and to $F>0$ for $g<0$. 

The last restriction \ arises from having \ considered the \ adiabatic speed of sound  $c_s^2=(\delta p/\delta\rho)_s=p_x/\rho_x$, (the subscript $s$ means at constant entropy), because the local stability and causality requirements $0 \le c_s^2 \le 1$ \cite{Hawking:1973uf,Wald:1984rg,Bean:2003fb,Longair:2008gba,Padmanabhan:1993aa,Coles:1995bd,Garriga:1999vw}
determine, through $c_s^2=F_x /(F_x +2xF_{xx})$ that the realistic models are those with $F_{xx}\le 0$. We use this last condition although in \cite{Babichev:2007dw} it is shown that condition $c_s^2 \le 1$ is not necessary for causality.

All the three conditions: $g<1$, $F_x>0$ and $F_{xx}\le 0$, are essential to describe  realistic models driven  by k-essence that are associated with acceptable interactions $ Q $ in the dark sector.\\

There are several functions that satisfy these three conditions. For example the quadratic function $ F [x] = -mx ^ 2 + nx + c $, which includes the linear one, the proportional to the tachyonic function \  $F [x] = m \sqrt{1 + x } + n $, the\  exponential \ $ F(x) = e^{-mx^2} + n $ \ and \ also \ $F(x) = -m \cosh(\sqrt{-x})$ with $ m > 0$, $ n > 0 $ and  $ c > 0 $. Some of them will be used in the next section to find the appropriate associated interaction $Q$ in the dark sector. 
The figure~\ref{fig:2} shows the EoS corresponding to functions  $F(x)_{stiff}= -mx+n$, $F(x)_{rad}= -mx^2+n$ and  $F(x)_{dust}= e^{-mx^2}+n$,  with $m>0$ and $n>0$, for which we can see the corresponding asymptotic limits.

Note that the global equation of state of the k-essence  is independent of the potential used and therefore the above results preserve their validity for variable forms of $V$, but in these last cases the first integral (\ref{12})  no longer exists. Moreover, note that the crossing of the phantom divide line (PDL) is not allowed. This was to the first time proven for k-essence in \cite{Vikman:2004dc}.


\section{The associate interactions}


The results of the previous section are quite general and apply to any kinetic function $F(x)$, but the particular choice of the function will be determined by the interaction $Q$ that manages the evolution of both fluids. 
The expressions (\ref{8b}) and (\ref{14}) let us write the equation that must be fulfilled by the kinetic function $F(x)$ once the interaction $Q_M(V_0,F)$ is fixed.
\be
\n{15}
\bigg(\frac{Q_M}{V_0}-B\frac{\left[A F - 2x(A-1)F_x\right]}{(A-B)}\bigg)\bigg(2M-(A-B)N\bigg) + 2xF_xAN=0,
\ee
\no with $ M=F_x+xF_{xx} $ and $ N=((2-A)F_x-2xF_{xx}(A-1))/(A-B) $. 

The expression $Q_M(V_0,F)$ means that the interaction, often expressed as a function of $\rho$  and its derivatives, should be given using equations (\ref{6}), (\ref{10}), (\ref{12}) and $\rho'= 2xF(x)V_0$.

The equation (\ref{15}) is a highly nonlinear equation for $F$. However, the change of variables $\zeta=\int {\rho_x/(2xF_xV_0)}dx$ and $\rho(x)=V_0(F-2xF_x)$  lets us to obtain the more simple  differential equation for $\rho$,  
\be
\n{16}
\rho'' + (A+B)\rho'+ AB\rho=Q_M(A-B)
\ee
\no with $\rho'=d\rho/d\zeta$ and $\rho''=d^2\rho/d\zeta^2$. This is the holographik version \cite{Chimento:2011dw} of the already known source equation for the energy density described in \cite{Chimento:2009hj}. On the other hand, equation (\ref{15}) allows using the representation in both directions. One direction is to find the system handled by the k essence $F$ that represents the interactive system and the other one is to assign an interaction $Q$ to an interactive system that is  studied as a  unified model of k essence. Let's have a look at some worked examples.
\vskip-1.5cm
\begin{itemize}     
\item
 \vskip0.5cm
\emph{Examples $Q \rightarrow F $}
 \vskip0.5cm
\begin{itemize}     
\item
\emph{CDM} and \emph{MHRDE}\\
This interesting case was already presented at the general formalism developed in \cite{Forte:2015oma}, where it was applied to the null interaction $Q=0$ or equivalently when we replace $Q_M=(1-A)\rho_1$ in (\ref{15}).  The solution is $F(x)=(F_0+F_1\sqrt{-x})^{B/(B-1)}$ with $F_0 < 0$, $F_1 > 0$, $0 < B < 1 $ and lets writing the densities of energy $\rho^{MHRDE}=b_1 a^{-3}+b_2 a^{-3B}$ and $\rho^{MHRDE}_2=((A-1)/(A-B))b_1 a^{-3}+b_2a^{-3B}$. It can be seen that the MHRDE fluid is always a self-interacting component, because even when $Q$ is null, the dark energy component is far from remaining independent of the CDM. The asymptotic values of the EoS $\omega=-b_2(1-B) a^{3(1-B)}/(b_1+b_2 a^{3(1-B)})$ are $ 0 $ and $B-1 < A-1$ in the asymptotic limits $ a \rightarrow 0 $ and $ a \rightarrow \infty $ respectively. However, the model is not viable because always $c_s^2 < 0$.

\vskip0.3cm
\item
\emph{The holographik $\Lambda$}\\

In this example we consider the case in which a holographic interactive fluid is behind the concept of cosmological constant.  The system of a CDM fluid interacting with a MHRDE fluid (\ref{5}) through the interaction $\Delta Q=B\rho+(1 - \Delta)\rho'$, $\Delta =A-B$,  can be interpreted as a cosmological model driven by a purely k-essence identified by the constant potential $V_0$ and the kinetic function obtained from (\ref{15}), \ \ $F=F_1 + 2F_0\frac{(1-A)}{A}(-x)^{\frac{A}{2(A-1)}}$,  \ \  with the positive constants of integration $F_0$ and $F_1$.
From  (\ref{10}), (\ref{12})  and (\ref{13b}) the expressions for the global density of energy and the density of energy of MHRDE fluid are 
\be
\n{17}
\rho=\Lambda \frac{A-B}{A} +\frac{\rho_{m}}{a^{3A}} \qquad  \rho_2=\Lambda,
\ee
\no respectively, with $\Lambda = AV_0F_1/(A-B)$ and $\rho_{m} =2V_0m_0^A/(AF_0^{A-1})$. 

 If $ 1<A<2 $, the corresponding global EoS
\be
\n{18}
\om= - \frac{\Lambda(A-B)+A(1-A)\rho_m a^{-3A}}{\Lambda(A-B)+A\rho_m a^{-3A}},
\ee
\no  ranges between the values $\om_{et}=A-1$ at early times and $\om_{lt}=-1$ at late times and the sound speed is $c_s^2=A-1 < 1$.

Solving  $\ro_2^{MHRDE}=(2\dot H +3 A H^2)/(A-B)=\Lambda$ we obtain the factor of scale 
\be
\n{19}
a(t)=\Big(\cosh(\kappa(t-t_0))+H_0 \sinh(\kappa(t-t_0))\Big)^{\frac{2}{3A}} \qquad \kappa^2=3A\Lambda(A-B)/4,
\ee 
\no where we set $t_0$ as the present time for which the factor of scale is $a(t_0)=1$ and Hubble parameter is $H(t_0)=H_0$. Notice that the argument of the hyperbolic functions in (\ref{19}) corresponds to the usual solution of the factor of scale for the model $\Lambda$CDM if $B=(A^2-1)/A$. Also, the expression (\ref{17}) corresponds to the model $\Lambda$ plus WDM ($A$ very slightly greater than one) or to the model $\Lambda$ plus radiation ($A=4/3$). However, unlike a true cosmological constant, the equation of state for dark energy $\om_2= \om \rho/\rho_2$ diverges at early times and tends asymptotically to $-\Delta/A$ at late times because its expression is
\be
\n{20}
\om_2= - \frac{(A-B)}{A}+\frac{(A-1)\rho_m }{\Lambda a^{3A}}.
\ee
\vskip0.3cm
\item
\emph{The sign-change holographik }\\

There exists a number of works that studied interactions able to change their sign along the evolution of the universe. One of them is $Q_{sc}=B\rho_2-\rho_1$ that replaced in (\ref{15}) allows us to obtain two linear differential equations $ xF_x - y^{\pm}F=0 $ where $ y^{\pm}=\sqrt{AB}/(2(\sqrt{AB}\pm 1)) $ and then, the two kinetic functions $F^{\pm}=F_0^{\pm}(-x)^{ y^{\pm}}$. Using the first integral (\ref{12}) for each particular kinetic function, the corresponding global energy density is
\be
\n{21}
\ro=(\ro_0-\ro_-)a^{3\sqrt{AB}} + \ro_-a^{-3\sqrt{AB}},\qquad\ro_-=\frac{V_0F^-_0}{(1-\sqrt{AB})},
\ee
\no with $\rho_0$ the actual global density, and the constant of integration $m_0^-$ coming from (\ref{12}) are taken so that $m_0^-=-F_0^-y^-$. Therefore, the global EoS oscillates between $-(1-\sqrt{AB})$ at early times and $-(1+\sqrt{AB})$ at late times as can be seen in 
\be
\n{22}
\om= - \frac{(\ro_0-\ro_-)(1+\sqrt{AB})a^{6\sqrt{AB}}+\ro_-(1-\sqrt{AB})}{(\ro_0-\ro_-)a^{6\sqrt{AB}}+\ro_-}.
\ee
 Assuming $\Delta>0$ and $\sqrt{AB}<1$, the change of sign of the interaction is produced at $\om_{sc}=(AB - 1)/(B+1)$ for which the factor of scale is
\be
\n{23}
a_{sc}=\Bigg(\frac{\ro_-(\sqrt{AB}(B+1)-(A+1)B)}{(\ro_0-\ro_-)(\sqrt{AB}(B+1)+(A+1)B)}\Bigg)^{\frac{1}{6\sqrt{AB}}}.
\ee
This interactive system affected by  $Q_{sc}$ is consistently maintained until $\rho_1$  is exhausted at 
\be
\n{24}
a_{max}= \Big(\frac{\ro_-(\sqrt{AB}-B)}{(\ro_0-\ro_-)(\sqrt{AB}+B)}\Big)^{\frac{1}{6\sqrt{AB}}},
\ee
\no when the sign change has already occurred because $a_{max}> a_{sc}$.
 
The figure~\ref{fig:3}  shows the global density of energy (\ref{21}) and the partial densities of energy
\begin{subequations}
\label{24}
\be
\label{24a} 
\ro_1=\frac{1}{\Delta} \Big\{ -(\ro_0-\ro_-)(\sqrt{AB}+B)a^{3\sqrt{AB}} + \ro_-(\sqrt{AB}-B)a^{-3\sqrt{AB}} \Big\},
\ee
\be
\label{24b} 
\ro_2=\frac{1}{\Delta}\Big\{(\ro_0-\ro_-)(A+\sqrt{AB})a^{3\sqrt{AB}} + \ro_-(A-\sqrt{AB})a^{-3\sqrt{AB}}\Big\},
\ee
\end{subequations}
where it can be seen that with the right choice of the constants of integration $(\ro_0-\ro_-)$ and $(\ro_-)$, the model is consistent with current estimates of dark energy densities and shows a relief in the problem of coincidence. Also there, the ratio $r=\rho_1/\rho_2$ and $Q_{sc}$ are depicted for $\ro_0-\ro_-=0.04$, $\ro_-=0.96$, $ A = 5/4$, $B = 3/4$.

\begin{figure}
\begin{center}
\includegraphics[height=8cm,width=10cm]{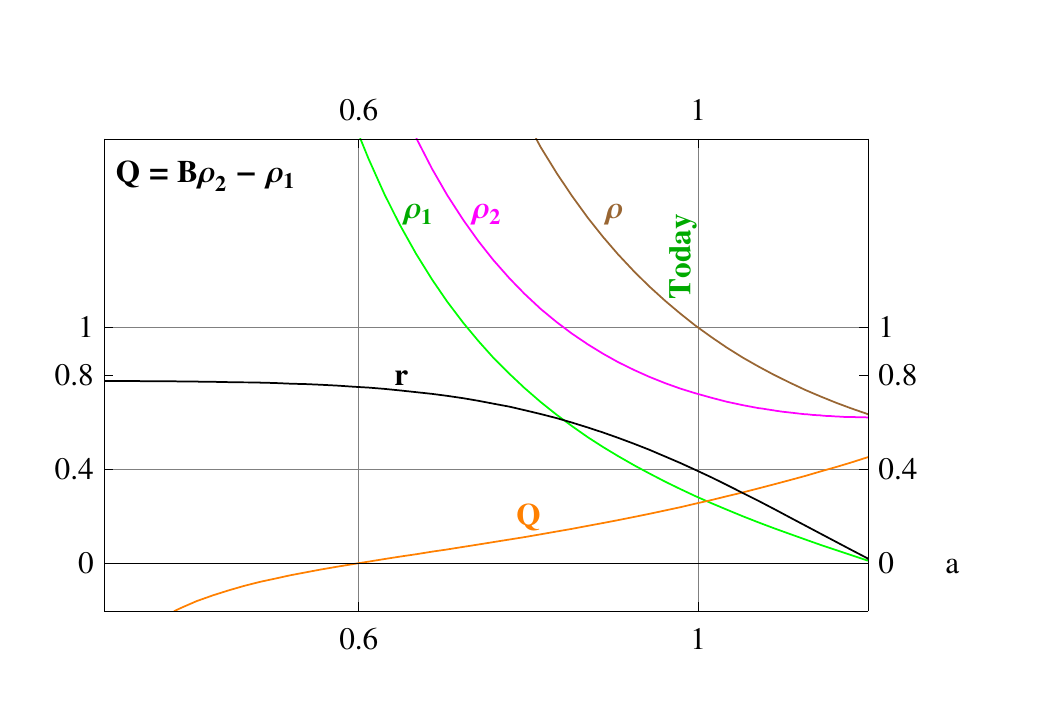}

\hfill

\caption{\label{fig:3} Densities of energy, ratio $r=\rho_1/\rho_2$ and $Q_{sc}$ for $\ro_0-\ro_-=0.04$, $\ro_-=0.96$, $ A = 5/4$, $B = 3/4$.}

\end{center}
\end{figure}

\end{itemize}   
 
\item
\vskip0.5cm
\emph{Examples $F \rightarrow Q $}
\begin{itemize} 
\item
\vskip0.5cm
The linear function $F(x)=1 + mx$, with $ m>0 $ was already used in 
\cite{Chimento:2003ta,Chimento:2005ua} with exponential potentials. Here, with the constant $V=V_0$, is replaced in (\ref{15}) obtaining the interaction $ Q_M\Delta = AB\rho + (A+B-2)\rho'$. That is, a cosmological model with CDM fluid interacting with MHRDE through the interaction $ \Delta Q = B\rho + (B-1)\rho'$ can be seen as a model driven by a purely linear k-essence. With the greatest simplicity, the densities of energy and the EoS of the  global model are obtained as well as the time variation of the scale of factor and the k-essence field $\phi$.

\begin{subequations}
\label{25}
\be
\label{25a}
\rho= V_0 + \frac{\rho^0}{a^6}, \qquad \rho_2= \frac{1}{\Delta}\big( AV_0 + (A-1)\frac{\rho^0}{a^6}\big),
\ee
\be
\label{25b}
a(t)=\Big[\frac{\sinh(\sqrt{3V_0}t)}{\sinh(\sqrt{3V_0}t_T)}\Big]^{1/3}\qquad \rho^0=\frac{m_0^2V_0}{m}, \qquad \omega= - \frac{V_0a^6 - \rho^0}{V_0a^6 + \rho^0},
\ee
\be
\label{25c}
\phi= \phi_0\ln(\tanh(\frac{\sqrt{3V_0} t}{2})) \qquad \phi_0=\frac{m_0\sinh(\sqrt{3V_0}t_T)}{m\sqrt{3V_0}},
\ee
\end{subequations}

\no with $ t_T $ the actual time.  The adiabatic velocity of sound is constantly equal to 1 as in the cases of the quintessence regardless of the values of the parameters $A$ and $B$.
\vskip0.5cm

\item
In \cite{Chimento:2004jm}, the simple quadratic function $F(x)=\frac{b}{6}+x-\frac{x^2}{2b}$ is used with the arbitrary parameter $b > 0$ to ensure positivity of density of energy and stability observed through the speed of sound. In this context of purely holographic k-essence with constant potential $V_0$ it leads, through (\ref{15}), to the associated interaction $Q_M\Delta=AB\rho+(A+B-1)\rho'+\rho''/4+ \rho'/(2\sqrt{6\rho/bV_0})$ or 
\be
\n{25}
Q\Delta=B\rho+B\rho'+\frac{\rho''}{4}+\frac{\rho'}{2\sqrt{\frac{6\rho}{bV_0}}}.
\ee

From (\ref{12}) we obtain the algebraic equation $h^3+h-u=0$ for $h=\sqrt{-x/b}$ and $u=m_0b^{-1/2}a^{-3}$, whose unique real solution allows us to write the first integral $\dot\phi=\sqrt{-x}$ for the kinetic function as
\be
\n{26}
\sqrt{\frac{-x}{b}}=-\frac{(2/3)^{1/3}b^{1/6}a}{\Big(9+\sqrt{3}\sqrt{4ba^6+27}\Big)^{1/3}}+\frac{\Big(9+\sqrt{3}\sqrt{4ba^6+27}\Big)^{1/3}}{2^{1/3}3^{2/3}b^{1/6}a},
\ee
\no and thus, from (\ref{10}) and (\ref{12}) the global energy density $\rho = \frac{3V_0b}{2}(1/3-x/b)^2$ proves to be 
\be
\n{27}
\rho=\frac{3V_0b}{2}\Bigg[\frac{1}{3}+\Bigg(-\frac{(2/3)^{1/3}b^{1/6}a}{\Big(9+\sqrt{3}\sqrt{4ba^6+27}\Big)^{1/3}}+\frac{\Big(9+\sqrt{3}\sqrt{4ba^6+27}\Big)^{1/3}}{2^{1/3}3^{2/3}b^{1/6}a}\Bigg)^2\Bigg]^2,
\ee
\no where the constant of integration in (\ref{12}) is taken as $m_0=1$.
The global EoS (\ref{14}), 
\be
\n{28}
\omega = -\frac{1}{3}\frac{(-\frac{x}{b}+1+\frac{2}{\sqrt{3}})(\frac{x}{b}-1+\frac{2}{\sqrt{3}})}{(-\frac{x}{b} +1/3)^2},
\ee
\no shows that this interactive model exhibits a dust type behavior $\omega=0$ when the time evolution of the k-essence is $\dot\phi=\sqrt{-x_{root}}= \sqrt{b(-1+2/\sqrt{3})}$, that is when $a_{dust}=1.3 b^{-1/6}$. The figure~\ref{fig:4} shows that the dust behavior of the global EoS can be accommodated by varying the parameter $b$ and also that the EoS has a single maximum regardless of $b$, corresponding to $A=3/2$.

\begin{figure}
\begin{center}
\includegraphics[height=8cm,width=10cm]{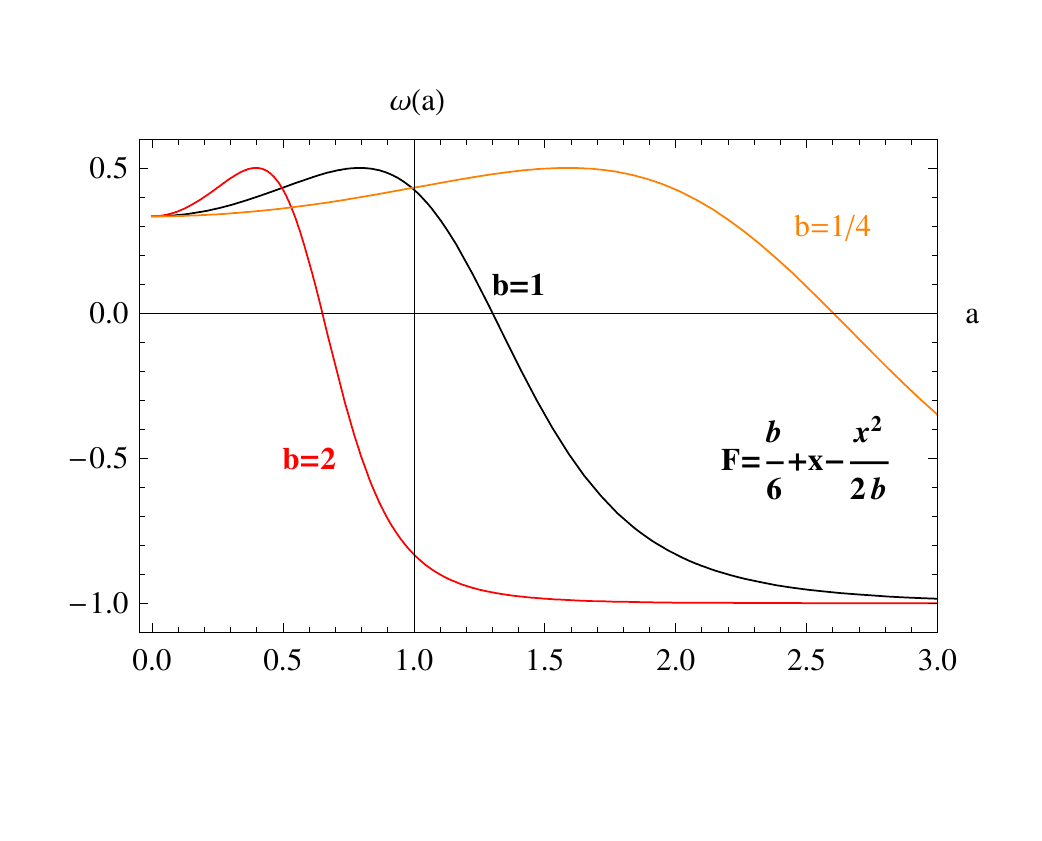}

\hfill

\caption{\label{fig:4} Global equation of state for the quadratic function $F(x)=\frac{b}{6}+x-\frac{x^2}{2b}$ for different parameters $b$. It can be seen that the maximum is independent of $b$}

\end{center}
\end{figure}

Thus, the constant $A$ is fixed for the kinetic function and $b$ is fixed by the astronomical data. The remaining constants $V_0$ and $B$ are determined by the current overall energy of density $\rho(a=1)$ and by the ratio between dark densities of energy $r=\rho_1/\rho_2=-(B(F-2xF_x)+2xF_x)/(AF-2xF_x(A-1))$ at its present value $r(a=1)$ respectively.

 The adiabatic velocity of sound $c_s^2=F_x/(F_x+2xF_xx)=(b-x)/(b-3x)$ oscillates between $c_s^2|_{et}=0$ at early time and $c_s^2|_{lt}=1/3$ at late times. 

\vskip0.5cm
\item
The kinetic function $F= \alpha - \beta \cosh(\sqrt{-x})$ with $\alpha>\beta>0$ meets the requirement $F_x= \beta \sinh(\sqrt{-x})/(2\sqrt{-x})>0$ in order that the density of energy is always positive and 
$F_{xx}= -\big(\beta\cosh(\sqrt{-x})/(4(-x)^{3/2}\big)\Big(\sqrt{-x}-\tanh(\sqrt{-x})\Big) < 0$ for the global model is stable. Then, using (\ref{12}) we find that $\sqrt{-x}=\sinh^{-1}(a^{-3})$ if the constant of integration is taken as $m_0=\beta/2$ and thus the global density of energy is

\be
\n{29}
\rho = V_0 \Big(\alpha-\frac{\beta}{a^3}\sqrt{1+a^6}+\frac{\beta}{a^3}\sinh^{-1}(a^{-3})\Big).
\ee

This expression (\ref{29}) can be considered as the global density of energy of an interactive cosmological system filled with CDM and MHRDE  that affected by the interaction $Q$
\be
\n{30}
\Delta Q = B(\rho+\rho')+\beta V_0\Big[\frac{\alpha V_0-(\rho+\rho')}{\beta V_0}-\frac{\beta V_0}{\alpha V_0-(\rho+\rho')}\Big],
\ee
\no produces a density of energy for the holographic component
\be
\n{31}
\rho_2^{MHRDE}= \frac{V_0}{\Delta}\Big[A\alpha - A\frac{\beta\sqrt{1+a^6}}{a^3}+\frac{\beta(A-1)}{a^3}\sinh^{-1}(a^{-3})\Big].
\ee
The corresponding global EoS 
\be
\n{32}
\omega =\frac{\alpha a^3 - \beta\sqrt{1+a^6}}{-\beta\sinh^{-1}(a^{-3})-\alpha a^3 + \beta\sqrt{1+a^6}},
\ee
\no has asymptotic values $\omega_e=0$ for $a=0$ and $\omega_l=-1$ at late times. Nevertheless the maximum value of $\omega$ is positive in the intermediate epoch between the asymptotic dust era $a=0$ and the truly dust era $a = \beta^{1/3}(\alpha-\beta)^{-1/6}(\alpha+\beta)^{-1/6} $  because 
$g= (\alpha-\beta\cosh(\sqrt{-x}))/(-2\beta\sqrt{-x}\sinh(\sqrt{-x})$ is not always negative at early times. Note that the constants $\alpha$ and $\beta$ should be adjusted so that the holographic density is kept positive and the quotient of densities in the dark sector fits with the current value. The adiabatic velocity of sound $c_s^2=(\sinh^{-1}(a^{-3})\sqrt{1+a^6})^{-1}$ oscillates between $c_s^2=0$ at early times and $c_s^2=1$ at late times, independently from the constants $A$, $B$, $\alpha$ and $\beta$.

\end{itemize}     
\end{itemize}



\section{Conclusions}

In this work we have studied cosmological models driven by k-essences with constant potential $V_0$, generated by strictly increasing ($F_x>0$) and concave ($F_{xx}<0$) kinetic functions $F$.  The study is described in a FRW background and the goal was the possibility of finding links between these universes and interactive models filled with CDM and MHRDE fluids.
This idea is supported on studies realized previously where scalar representations of cosmological interactive arbitrary systems were found using exotic quintessence \cite{Chimento:2007da} and exotic k-essence as scalar fields \cite{Forte:2015oma}.
Here we particularize the interactive model, considering it as integrated by CDM and MHRDE fluid, whose defining parameters $A$ and $B$ mark limits on the used k-essences. According to the general method described in \cite{Forte:2015oma}, the fields turn out to be common k-essences derived from a Lagrangian $\mathcal{L}=-V_0F$. This fact could allow us to consider this formalism as an indirect or covert Lagrangian description of a cosmological system with interactive dark energy \cite{Poplawski:2006ey}.
In the k-essential approach of the cosmological models they do not have allowed the crossing of phantom divide ($\om=-1$). This is clearly seen in the figure~\ref{fig:1} where the global EoS $\om = g/(1-g)$ is plotted as function of an auxiliary magnitude $g=F/2xF_x$. There are two branches in the picture, one describing viable universes and the other corresponding to phantom universes, and the cosmological constant type behavior is the asymptotic conduct in the extreme points ($F_x=0$), when we arrive at the low limit $g\rightarrow-\infty$ of the acceptable branch $g < 1$ or at the top limit $g\rightarrow + \infty$ of the phantom branch $g > 1$.
Restricting ourselves to models no phantom for which $g < 1$, a positive global density of energy leads to the condition $F_x>0$ while the stability measured by the adiabatic speed of sound determines $F_{xx}<0$. 
These conditions select possible k-essences for the representation ($F \rightarrow Q$) and simultaneously they reject interactions that can lead, through this approach, to cosmologically nonviable systems ($Q \rightarrow F$). 
Two other restrictions on these models of universe arising from the use of $CDM$ and $MHRDE$ as interacting fluids are the inability to have constant ratios $r=r_0$ to alleviate the problem of the coincidence without having a global constant EoS $\om=\om_0$ (because $ r = r_0 =(A-1-\om_0)/(1+\om_0-B)$), and the existence of a maximum value for the global EoS $\om_{max}=A-1$.

The link between both schemes arises from equalizing the expressions of the time derivative of Hubble parameter $-2\dot H = (\rho + p)= -2xF_xV_0 $ in each description and from supposing a linear combination of density of energy and of pressure for the $DE$. Then, the conservation equation for the $DE$ gives us the expression that must be hold for the kinetic function $F$ and for the interaction $Q$. From (\ref{15}), given the interaction $Q(V_0,F)$, the finding of the function $F$ allows to write all the densities of energy as function of the factor of scale through the expressions (\ref{10}), (\ref{12}) and (\ref{13}). Inversely, given the appropriate function $F$ we find the interaction that affects the CDM-MHRDE system. The last two sections were dedicated to giving examples of these two manners of using (\ref{15}). To the systems CDM-MHRDE affected by the interactions $Q=0$, $Q_{\Lambda}=(B\rho+(1 - \Delta)\rho')/\Delta $,  and $Q_{sc}=B\rho_2-\rho_1$  we associate systems driven by the k-essences $F(x)=(F_0+F_1\sqrt{-x})^{B/(B-1)}$, $F=F_1 + 2F_0\frac{(1-A)}{A}(-x)^{\frac{A}{2(A-1)}}$ and $F^{\pm}=F_0^{\pm}(-x)^{ y^{\pm}}$ respectively, showing that they arrive at the same dynamic results in both approaches, but in a more direct way. Also, in the particular case $Q_{\Lambda}$ we show that the concept of cosmological constant can be interpreted as the result of an interaction in these systems CDM-MHRDE.
For the inverse way, we use the kinetic functions $F(x)=1+mx$, $F(x)=\frac{b}{6}+x-\frac{x^2}{2b}$ and $F= \alpha - \beta \cosh(\sqrt{-x})$ to obtain interactions not usually considered in the literature, $ Q = (B\rho + (B-1)\rho')/\Delta $, $Q=(B\rho+B\rho'+\frac{\rho''}{4}+\frac{\rho'(6\rho/bV_0)^{-1/2}}{2})/\Delta $,  and $Q = [B(\rho+\rho')+\beta V_0[\frac{\alpha V_0-(\rho+\rho')}{\beta V_0}-\frac{\beta V_0}{\alpha V_0-(\rho+\rho')}]]/\Delta $ respectively. The latter case shows the difference between the maximum value of the global EoS and its asymptotic limits. Moreover in the last two cases, we obtained interactions that difficultly could be solved by the method of the equation source and for which, nevertheless, we obtained explicit expressions for all the cosmological important magnitudes.



\begin{thebibliography}{99}


\bibitem{Riess:1998cb} 
  A.~G.~Riess {\it et al.} [Supernova Search Team Collaboration],
  Astron.\ J.\  {\bf 116}, 1009 (1998)
  doi:10.1086/300499
  [astro-ph/9805201].

\bibitem{Perlmutter:1998np} 
  S.~Perlmutter {\it et al.} [Supernova Cosmology Project Collaboration],
  Astrophys.\ J.\  {\bf 517}, 565 (1999)
  doi:10.1086/307221
  [astro-ph/9812133].

\bibitem{Amanullah:2010vv} 
  R.~Amanullah {\it et al.},
  Astrophys.\ J.\  {\bf 716}, 712 (2010)
  doi:10.1088/0004-637X/716/1/712
  [arXiv:1004.1711 [astro-ph.CO]].

\bibitem{Spergel:2006hy} 
  D.~N.~Spergel {\it et al.} [WMAP Collaboration],
  Astrophys.\ J.\ Suppl.\  {\bf 170}, 377 (2007)
  doi:10.1086/513700
  [astro-ph/0603449].

\bibitem{Weinberg:2012es} 
  D.~H.~Weinberg, M.~J.~Mortonson, D.~J.~Eisenstein, C.~Hirata, A.~G.~Riess and E.~Rozo,
  Phys.\ Rept.\  {\bf 530}, 87 (2013)
  doi:10.1016/j.physrep.2013.05.001
  [arXiv:1201.2434 [astro-ph.CO]].

\bibitem{Percival:2007yw} 
  W.~J.~Percival, S.~Cole, D.~J.~Eisenstein, R.~C.~Nichol, J.~A.~Peacock, A.~C.~Pope and A.~S.~Szalay,
  Mon.\ Not.\ Roy.\ Astron.\ Soc.\  {\bf 381}, 1053 (2007)
  doi:10.1111/j.1365-2966.2007.12268.x
  [arXiv:0705.3323 [astro-ph]].

\bibitem{Zlatev:1998tr} 
  I.~Zlatev, L.~M.~Wang and P.~J.~Steinhardt,
  Phys.\ Rev.\ Lett.\  {\bf 82}, 896 (1999)
  doi:10.1103/PhysRevLett.82.896
  [astro-ph/9807002].
\bibitem{Caldwell:1997ii}
  R.~R.~Caldwell, R.~Dave and P.~J.~Steinhardt,
  Phys.\ Rev.\ Lett.\  {\bf 80} (1998) 1582
  doi:10.1103/PhysRevLett.80.1582
  [astro-ph/9708069].
\bibitem{Frieman:1995pm} 
  J.~A.~Frieman, C.~T.~Hill, A.~Stebbins and I.~Waga,
  Phys.\ Rev.\ Lett.\  {\bf 75}, 2077 (1995)
  doi:10.1103/PhysRevLett.75.2077
  [astro-ph/9505060].
\bibitem{Peebles:1987ek} 
  P.~J.~E.~Peebles and B.~Ratra,
  Astrophys.\ J.\  {\bf 325}, L17 (1988).
  doi:10.1086/185100
\bibitem{Ratra:1987rm} 
  B.~Ratra and P.~J.~E.~Peebles,
  Phys.\ Rev.\ D {\bf 37}, 3406 (1988).
  doi:10.1103/PhysRevD.37.3406



\bibitem{ArmendarizPicon:2000dh} 
  C.~Armendariz-Picon, V.~F.~Mukhanov and P.~J.~Steinhardt,
  Phys.\ Rev.\ Lett.\  {\bf 85}, 4438 (2000)
  doi:10.1103/PhysRevLett.85.4438
  [astro-ph/0004134].
\bibitem{ArmendarizPicon:2000ah} 
  C.~Armendariz-Picon, V.~F.~Mukhanov and P.~J.~Steinhardt,
  Phys.\ Rev.\ D {\bf 63}, 103510 (2001)
  doi:10.1103/PhysRevD.63.103510
  [astro-ph/0006373].
\bibitem{Copeland:2006wr} 
  E.~J.~Copeland, M.~Sami and S.~Tsujikawa,
  Int.\ J.\ Mod.\ Phys.\ D {\bf 15}, 1753 (2006)
  doi:10.1142/S021827180600942X
  [hep-th/0603057].
\bibitem{Chimento:2003ta} 
  L.~P.~Chimento,
  Phys.\ Rev.\ D {\bf 69}, 123517 (2004)
  doi:10.1103/PhysRevD.69.123517
  [astro-ph/0311613].
\bibitem{Chiba:2002mw} 
  T.~Chiba,
  Phys.\ Rev.\ D {\bf 66}, 063514 (2002)
  doi:10.1103/PhysRevD.66.063514
  [astro-ph/0206298].
\bibitem{Tsujikawa:2004dp} 
  S.~Tsujikawa and M.~Sami,
  Phys.\ Lett.\ B {\bf 603}, 113 (2004)
  doi:10.1016/j.physletb.2004.10.023
  [hep-th/0409212].
\bibitem{Malquarti:2003nn} 
  M.~Malquarti, E.~J.~Copeland, A.~R.~Liddle and M.~Trodden,
  Phys.\ Rev.\ D {\bf 67}, 123503 (2003)
  doi:10.1103/PhysRevD.67.123503
  [astro-ph/0302279].
\bibitem{Aguirregabiria:2004te} 
  J.~M.~Aguirregabiria, L.~P.~Chimento and R.~Lazkoz,
  Phys.\ Rev.\ D {\bf 70}, 023509 (2004)
  doi:10.1103/PhysRevD.70.023509
  [astro-ph/0403157].
\bibitem{Chimento:2004jm} 
  L.~P.~Chimento, M.~I.~Forte and R.~Lazkoz,
  Mod.\ Phys.\ Lett.\ A {\bf 20}, 2075 (2005)
  doi:10.1142/S0217732305018074
  [astro-ph/0407288].
\bibitem{Chimento:2005ua} 
  L.~P.~Chimento and M.~I.~Forte,
  Phys.\ Rev.\ D {\bf 73}, 063502 (2006)
  doi:10.1103/PhysRevD.73.063502
  [astro-ph/0510726].
\bibitem{ArmendarizPicon:2005nz} 
  C.~Armendariz-Picon and E.~A.~Lim,
  JCAP {\bf 0508}, 007 (2005)
  doi:10.1088/1475-7516/2005/08/007
  [astro-ph/0505207].
\bibitem{Li:2002wd} 
  M.~Li and X.~Zhang,
  Phys.\ Lett.\ B {\bf 573}, 20 (2003)
  doi:10.1016/j.physletb.2003.08.041
  [hep-ph/0209093].










\bibitem{Guo:2004fq} 
  Z.~K.~Guo, Y.~S.~Piao, X.~M.~Zhang and Y.~Z.~Zhang,
  Phys.\ Lett.\ B {\bf 608}, 177 (2005)
  doi:10.1016/j.physletb.2005.01.017
  [astro-ph/0410654].
\bibitem{Cai:2009zp} 
  Y.~F.~Cai, E.~N.~Saridakis, M.~R.~Setare and J.~Q.~Xia,
  Phys.\ Rept.\  {\bf 493}, 1 (2010)
  doi:10.1016/j.physrep.2010.04.001
  [arXiv:0909.2776 [hep-th]].
\bibitem{Feng:2004ad} 
  B.~Feng, X.~L.~Wang and X.~M.~Zhang,
  Phys.\ Lett.\ B {\bf 607}, 35 (2005)
  doi:10.1016/j.physletb.2004.12.071
  [astro-ph/0404224].
\bibitem{Feng:2004ff} 
  B.~Feng, M.~Li, Y.~S.~Piao and X.~Zhang,
  Phys.\ Lett.\ B {\bf 634}, 101 (2006)
  doi:10.1016/j.physletb.2006.01.066
  [astro-ph/0407432].
\bibitem{Chimento:2008ws} 
  L.~P.~Chimento, M.~I.~Forte, R.~Lazkoz and M.~G.~Richarte,
  Phys.\ Rev.\ D {\bf 79}, 043502 (2009)
  doi:10.1103/PhysRevD.79.043502
  [arXiv:0811.3643 [astro-ph]].

\bibitem{'tHooft:1993gx} 
  G.~'t Hooft,
  Salamfest 1993:0284-296
  [gr-qc/9310026].

\bibitem{Fischler:1998st} 
  W.~Fischler and L.~Susskind,
[hep-th/9806039].

\bibitem{Aharony:1999ti} 
  O.~Aharony, S.~S.~Gubser, J.~M.~Maldacena, H.~Ooguri and Y.~Oz,
  Phys.\ Rept.\  {\bf 323}, 183 (2000)
  doi:10.1016/S0370-1573(99)00083-6
  [hep-th/9905111].

\bibitem{Bousso:2002ju} 
  R.~Bousso,
  Rev.\ Mod.\ Phys.\  {\bf 74}, 825 (2002)
  doi:10.1103/RevModPhys.74.825
  [hep-th/0203101].
\bibitem{Bousso:1999xy} 
  R.~Bousso,
  JHEP {\bf 9907}, 004 (1999)
  doi:10.1088/1126-6708/1999/07/004
  [hep-th/9905177].
\bibitem{Hsu:2004ri} 
  S.~D.~H.~Hsu,
  Phys.\ Lett.\ B {\bf 594}, 13 (2004)
  doi:10.1016/j.physletb.2004.05.020
  [hep-th/0403052].





\bibitem{Amendola:1999er} 
  L.~Amendola,
  Phys.\ Rev.\ D {\bf 62}, 043511 (2000)
  doi:10.1103/PhysRevD.62.043511
  [astro-ph/9908023].
\bibitem{Mangano:2002gg} 
  G.~Mangano, G.~Miele and V.~Pettorino,
  Mod.\ Phys.\ Lett.\ A {\bf 18}, 831 (2003)
  doi:10.1142/S0217732303009940
  [astro-ph/0212518].
\bibitem{delCampo:2004wc} 
  S.~del Campo, R.~Herrera and D.~Pavon,
  Phys.\ Rev.\ D {\bf 70}, 043540 (2004)
  doi:10.1103/PhysRevD.70.043540
  [astro-ph/0407047].
\bibitem{Farrar:2003uw} 
  G.~R.~Farrar and P.~J.~E.~Peebles,
  Astrophys.\ J.\  {\bf 604}, 1 (2004)
  doi:10.1086/381728
  [astro-ph/0307316].
\bibitem{Guo:2004xx} 
  Z.~K.~Guo, R.~G.~Cai and Y.~Z.~Zhang,
  JCAP {\bf 0505}, 002 (2005)
  doi:10.1088/1475-7516/2005/05/002
  [astro-ph/0412624].
\bibitem{Guo:2004vg} 
  Z.~K.~Guo and Y.~Z.~Zhang,
  Phys.\ Rev.\ D {\bf 71}, 023501 (2005)
  doi:10.1103/PhysRevD.71.023501
  [astro-ph/0411524].
\bibitem{Cai:2004dk} 
  R.~G.~Cai and A.~Wang,
  JCAP {\bf 0503}, 002 (2005)
  doi:10.1088/1475-7516/2005/03/002
  [hep-th/0411025].
\bibitem{Gumjudpai:2005ry} 
  B.~Gumjudpai, T.~Naskar, M.~Sami and S.~Tsujikawa,
  JCAP {\bf 0506}, 007 (2005)
  doi:10.1088/1475-7516/2005/06/007
  [hep-th/0502191].
\bibitem{Curbelo:2005dh} 
  R.~Curbelo, T.~Gonzalez, G.~Leon and I.~Quiros,
  Class.\ Quant.\ Grav.\  {\bf 23}, 1585 (2006)
  doi:10.1088/0264-9381/23/5/010
  [astro-ph/0502141].

\bibitem{Zimdahl:2001ar} 
  W.~Zimdahl and D.~Pavon,
  Phys.\ Lett.\ B {\bf 521}, 133 (2001)
  doi:10.1016/S0370-2693(01)01174-1
  [astro-ph/0105479].

\bibitem{Chimento:2003iea} 
  L.~P.~Chimento, A.~S.~Jakubi, D.~Pavon and W.~Zimdahl,
  Phys.\ Rev.\ D {\bf 67}, 083513 (2003)
  doi:10.1103/PhysRevD.67.083513
  [astro-ph/0303145].


\bibitem{Olivares:2005tb} 
  G.~Olivares, F.~Atrio-Barandela and D.~Pavon,
  Phys.\ Rev.\ D {\bf 71}, 063523 (2005)
  doi:10.1103/PhysRevD.71.063523
  [astro-ph/0503242].


\bibitem{Szydlowski:2005ph} 
 M.~Szydlowski,
``Cosmological model with energy transfer,''
 Phys.\ Lett.\ B {\bf 632}, 1 (2006)
 doi:10.1016/j.physletb.2005.10.039
[astro-ph/0502034].
\bibitem{Chimento:2007da} 
  L.~Chimento and M.~I.~Forte,
  Phys.\ Lett.\ B {\bf 666}, 205 (2008)
  doi:10.1016/j.physletb.2008.07.064
  [arXiv:0706.4142 [astro-ph]].
\bibitem{Forte:2016cwm} 
  M.~Forte,
  arXiv:1610.07441 [hep-th].

\bibitem{Forte:2015oma} 
  M.~Forte,
  Eur.\ Phys.\ J.\ C {\bf 76}, no. 1, 42 (2016)
  doi:10.1140/epjc/s10052-016-3882-6
  [arXiv:1507.03658 [gr-qc]].















\bibitem{Granda:2008tm} 
  L.~N.~Granda and A.~Oliveros,
  Phys.\ Lett.\ B {\bf 671}, 199 (2009)
  doi:10.1016/j.physletb.2008.12.025
  [arXiv:0810.3663 [gr-qc]].
\bibitem{Nojiri:2005pu} 
  S.~Nojiri and S.~D.~Odintsov,
  Gen.\ Rel.\ Grav.\  {\bf 38}, 1285 (2006)
  doi:10.1007/s10714-006-0301-6
  [hep-th/0506212].

\bibitem{Li:2004rb} 
  M.~Li,
  Phys.\ Lett.\ B {\bf 603}, 1 (2004)
  doi:10.1016/j.physletb.2004.10.014
  [hep-th/0403127].
  




\bibitem{Chimento:2011dw} 
  L.~P.~Chimento, M.~I.~Forte and M.~G.~Richarte,
  Mod.\ Phys.\ Lett.\ A {\bf 28}, 1250235 (2013)
  doi:10.1142/S0217732312502355
  [arXiv:1106.0781 [astro-ph.CO]].
\bibitem{Chimento:2013se} 
  L.~P.~Chimento, M.~Forte and M.~G.~Richarte,
  Eur.\ Phys.\ J.\ C {\bf 73}, no. 1, 2285 (2013)
  doi:10.1140/epjc/s10052-013-2285-1
  [arXiv:1301.2737 [gr-qc]].

\bibitem{P:2013cmq} 
  P.~Pankunni and T.~K.~Mathew,
  Int.\ J.\ Mod.\ Phys.\ D {\bf 23}, 1450024 (2014)
  doi:10.1142/S0218271814500242
  [arXiv:1309.3136 [astro-ph.CO]].

\bibitem{Li:2014eba} 
  E.~K.~Li, Y.~Zhang and J.~L.~Geng,
  Phys.\ Rev.\ D {\bf 90}, no. 8, 083534 (2014)
  doi:10.1103/PhysRevD.90.083534
  [arXiv:1412.5482 [gr-qc]].

\bibitem{Pasqua:2015bpm} 
  A.~Pasqua, S.~Chattopadhyay and R.~Myrzakulov,
  arXiv:1511.00600 [gr-qc].

\bibitem{Sharif:2012zza} 
  M.~Sharif and A.~Jawad,
  Astrophys.\ Space Sci.\  {\bf 337}, 789 (2012).
  doi:10.1007/s10509-011-0893-5

\bibitem{delCampo:2011jp} 
  S.~del Campo, J.~C.~Fabris, R.~Herrera and W.~Zimdahl,
  Phys.\ Rev.\ D {\bf 83}, 123006 (2011)
  doi:10.1103/PhysRevD.83.123006
  [arXiv:1103.3441 [astro-ph.CO]].
 
\bibitem{Landim:2015hqa} 
  R.~C.~G.~Landim,
  Int.\ J.\ Mod.\ Phys.\ D {\bf 25}, no. 04, 1650050 (2016)
  doi:10.1142/S0218271816500504
  [arXiv:1508.07248 [hep-th]].

\bibitem{Bamba:2012cp} 
  K.~Bamba, S.~Capozziello, S.~Nojiri and S.~D.~Odintsov,
  Astrophys.\ Space Sci.\  {\bf 342}, 155 (2012)
  doi:10.1007/s10509-012-1181-8
  [arXiv:1205.3421 [gr-qc]].



\bibitem{Granda:2008dk} 
  L.~N.~Granda and A.~Oliveros,
  Phys.\ Lett.\ B {\bf 669}, 275 (2008)
  doi:10.1016/j.physletb.2008.10.017
  [arXiv:0810.3149 [gr-qc]].
 


\bibitem{Zeldovich:1972zz} 
  Y.~B.~Zeldovich,
  Mon.\ Not.\ Roy.\ Astron.\ Soc.\  {\bf 160}, 1P (1972).

\bibitem{Chavanis:2014lra} 
  P.~H.~Chavanis,
  Phys.\ Rev.\ D {\bf 92}, no. 10, 103004 (2015)
  doi:10.1103/PhysRevD.92.103004
  [arXiv:1412.0743 [gr-qc]].




\bibitem{Hawking:1973uf} 
  S.~W.~Hawking and G.~F.~R.~Ellis,
  doi:10.1017/CBO9780511524646
\bibitem{Wald:1984rg} 
  R.~M.~Wald,
  Chicago, Usa: Univ. Pr. ( 1984) 491p
  doi:10.7208/chicago/9780226870373.001.0001

\bibitem{Bean:2003fb} 
  R.~Bean and O.~Dore,
  Phys.\ Rev.\ D {\bf 69}, 083503 (2004)
  doi:10.1103/PhysRevD.69.083503
  [astro-ph/0307100].
\bibitem{Longair:2008gba} 
  M.~S.~Longair,
  doi:10.1007/978-3-540-73478-9


\bibitem{Padmanabhan:1993aa}
  ~T.~Padmanabhan,
   Cambridge: Cambridge University Press. (1993)

\bibitem{Coles:1995bd} 
  P.~Coles and F.~Lucchin,
  Chichester, UK: Wiley (2002) 492 p

\bibitem{Garriga:1999vw} 
  J.~Garriga and V.~F.~Mukhanov,
  Phys.\ Lett.\ B {\bf 458}, 219 (1999)
  doi:10.1016/S0370-2693(99)00602-4
  [hep-th/9904176].




\bibitem{Babichev:2007dw} 
  E.~Babichev, V.~Mukhanov and A.~Vikman,
  JHEP {\bf 0802}, 101 (2008)
  doi:10.1088/1126-6708/2008/02/101
  [arXiv:0708.0561 [hep-th]].
\bibitem{Vikman:2004dc} 
  A.~Vikman,
  Phys.\ Rev.\ D {\bf 71}, 023515 (2005)
  doi:10.1103/PhysRevD.71.023515
  [astro-ph/0407107].


\bibitem{Chimento:2009hj} 
  L.~P.~Chimento,
  Phys.\ Rev.\ D {\bf 81}, 043525 (2010)
  doi:10.1103/PhysRevD.81.043525
  [arXiv:0911.5687 [astro-ph.CO]].


\bibitem{Poplawski:2006ey} 
  N.~J.~Poplawski,
  gr-qc/0608031.


\end{thebibliography}
\end{document}